\documentclass[twocolumn,prl,superscriptaddress,nofootinbib]{revtex4-1}
\usepackage[english]{babel}
\usepackage[ansinew]{inputenc}
\usepackage{boldline,multirow,xcolor,colortbl}
\usepackage{times}
\usepackage{graphicx}
\usepackage{graphics}
\usepackage{amsmath}
\usepackage{amsfonts}
\usepackage{amssymb}
\usepackage{amsbsy}
\usepackage{dsfont}
\usepackage{epstopdf}
\usepackage{makeidx}
\usepackage{subfigure}
\usepackage{color} 
\usepackage{pgf}
\usepackage{bm}
\usepackage{tikz} 
\usepackage[normalem]{ulem}
\newcommand{\be}{\begin{equation}}
\newcommand{\ee}{\end{equation}}
\newcommand{\bea}{\begin{eqnarray}}
\newcommand{\eea}{\end{eqnarray}}

\setcitestyle{super,open={},close={}}

\makeindex
\begin{document}

\title{Two-photon spontaneous emission in atomically thin plasmonic nanostructures}

\author{Y. Muniz}
\affiliation{Instituto  de  F\'isica,  Universidade  Federal  do Rio  de  Janeiro,  Caixa  Postal  68528,   Rio  de  Janeiro  21941-972,  RJ,  Brazil}
\affiliation{Theoretical Division, Los Alamos National Laboratory, Los Alamos, NM 87545, USA}
\affiliation{Center for Nonlinear Studies, Los Alamos National Laboratory, Los Alamos, NM 87545,USA}

\author{A. Manjavacas}
\affiliation{Department of Physics and Astronomy, University of New Mexico, Albuquerque, NM 87131, USA
}

\author{C. Farina}
\affiliation{Instituto  de  F\'isica,  Universidade  Federal  do Rio  de  Janeiro,  Caixa  Postal  68528,   Rio  de  Janeiro  21941-972,  RJ,  Brazil}

\author{D. A. R. Dalvit}
\affiliation{Theoretical Division,  Los Alamos National Laboratory,  Los Alamos, NM 87545, USA}

\author{W. J. M. Kort-Kamp$^{*,}$}
\affiliation{Theoretical Division, Los Alamos National Laboratory, Los Alamos, NM 87545, USA}

\begin{abstract}
The ability to harness light-matter interactions at the few-photon level plays a pivotal role in quantum technologies. Single photons - the most elementary states of light - can be generated on-demand in atomic and solid state emitters. Two-photon states are also key quantum assets, but achieving them in individual emitters is challenging because their generation rate  is much slower than competing one-photon processes. We demonstrate that atomically thin plasmonic nanostructures can harness two-photon spontaneous emission, resulting in giant far-field two-photon production, a wealth of resonant modes enabling tailored photonic and plasmonic entangled states, and plasmon-assisted single-photon creation orders of magnitude more efficient than standard one-photon emission. We unravel the two-photon spontaneous emission channels and show  that their spectral line-shapes emerge from an intricate interplay between Fano and Lorentzian resonances. Enhanced two-photon spontaneous emission in two-dimensional nanostructures paves the way to an alternative efficient source of light-matter entanglement for on-chip quantum information processing and free-space quantum communications.
\end{abstract}

\maketitle

The generation of non-classical states of light has become a sought-after goal in nanophotonics in recent years, including  production of single photons  from atomic~\cite{haroche2013} and solid state~\cite{aharonovich2016} emitters on-demand, and entangled photon pairs 
in nonlinear crystals~\cite{kwiat95}. {\it Two-photon} spontaneous emission (TPSE) processes~\cite{goppert1931, zalialiutdinov2018} can also generate entangled photons and have been demonstrated in atomic~\cite{lipeles1965, bannett1982, cesar1996}, semiconductor~\cite{hayat2008}, and biexciton-exciton decay in quantum dots~\cite{ota2011, wang2019}. 
Nevertheless, the TPSE rate is typically eight to five orders of magnitude slower than competing one-photon decay rates. Intense plasmonic electromagnetic fields are known to enhance light emission via the Purcell effect~\cite{Tame2013,lodahl2015, poddubny2012}, and plasmon-assisted collective TPSE has been measured in bulk semiconductors coupled to nanoantenna arrays~\cite{nevet2010}  with only a  few tens of radiative emission enhancement.  On the other hand, spontaneous decay into {\it two-plasmon polaritons}  in bulk metals~\cite{goncalves2020}  and graphene monolayers~\cite{rivera2016} is predicted to be more than ten orders of magnitude larger than two-photon transitions. Polar dielectrics have also been proposed to enable  {\it two-phonon polariton} emission faster than competing single-phonon processes~\cite{rivera2017}. However, these conventional surface wave polaritons yield a rather simple broadband emission spectrum, are intrinsically non-radiative, and out-coupling them into far-field radiation by, {\it e.g.},  defect engineering, while maintaining a high Purcell factor is challenging~\cite{hoang2015} and generally leads to inefficient photon production.

Here, we show that two-dimensional plasmonic nanostructures are an ideal material platform to harness two-quanta emission processes from single emitters~\cite{note1}, enabling emission rates significantly faster than in monolayers and thin films. We develop a comprehensive study of the dominant two-quanta decay channels in finite-sized ultra-thin structures with arbitrary shape and material properties, unravelling an intricate interplay of Fano and Lorentzian lineshapes in single, dual, and even multiband emission. We report giant emission of photon-pairs enabled by localized surface plasmons supported in 2D nanostructures, which naturally leak into photonic modes and result in radiative TPSE several orders of magnitude larger than via ordinary surface plasmon polaritons. We discover a surprising TPSE effect arising from the existence of dark plasmonic modes in finite-sized 2D plasmonic nanostructures, which make  photon production through two-quanta transitions to be more efficient than via standard one-photon processes. Finally, we argue that our findings can be experimentally verified using recent advances in fabrication of ultra-thin plasmonic nanostructures~\cite{maniyara2019, el-fattah2019}, and state-of-the-art photo-coincidence~\cite{hayat2008, wang2019} and time-resolved fluorescence spectroscopy~\cite{lodhal2004,hoang2015} techniques.

Let us consider a quantum emitter placed in the proximity of an arbitrarily shaped 2D nanostructure (Fig. 1), and study its decay from an initial state of energy $\hbar\omega_i$ to a final one $\hbar\omega_f$ via two-quanta processes assisted by intermediate states of energy $\hbar\omega_m$. The TPSE rate for an emitter at position ${\bf R}_e$ with size $l_e$ much smaller than the transition wavelengths is~\cite{muniz2019, SI}
\begin{equation}
\Gamma({\bf R}_e)\! =\!\!  \int_0^{\omega_t}\!\!\!\!\!\!\! d\omega \gamma_0(\omega)\!\!\sum_{a,b}\!t_{ab}(\omega)
P_a({\bf R}_e,\omega)P_b({\bf R}_e,\omega_t-\omega),\!
\label{GammaTotal}
\end{equation}
where 
$\omega_t = \omega_i-\omega_f$ is the transition frequency,
$\gamma_0(\omega)\sim \omega^3(\omega_t-\omega)^3 l_e^6/c^6$ is the free-space TPSE spectral density~\cite{craig84}, 
and $t_{ab}(\omega)$ is a tensor that depends only on the electronic structure of the emitter.
$P_a({\bf R}_e,\omega)$ is the Purcell factor for a transition dipole moment oriented along the direction of the unit vector ${\bf \hat{e}}_a$ ($a = 1, 2, 3$)~\cite{note2,novotny2012,milonni2013},
and is proportional to the local density of states~\cite{carminati2015} that can be tailored with properly designed photonic environments. The TPSE spectrum $\gamma({\bf R}_e,\omega)$, {\em i.e.}, the integrand in Eq. \eqref{GammaTotal}, is always symmetric with respect to $\omega_t/2$ due to energy conservation. Two-quanta spontaneous emission close to low-loss 
plasmonic media is mainly driven by three unique photon-photon, photon-plasmon, and plasmon-plasmon relaxation channels.  The contribution of these pathways to the TPSE can be computed through the decomposition of the Purcell factors into their radiative and non-radiative parts~\cite{novotny2012}, $P_a(\omega) = P_{a,r}(\omega)+ P_{a,nr}(\omega)$ (from now on the ${\bf R}_e$ dependency is implict). For example, the spectral photon-photon TPSE  rate is given by 
\begin{equation}
\gamma_{ph,ph}(\omega)\! =\gamma_0(\omega)\!\sum_{a,b}t_{ab}(\omega)
P_{a,r}(\omega)P_{b,r}(\omega_t-\omega),
\label{gammaphph}
\end{equation}
and similar expressions hold for the photon-plasmon $\gamma_{ph,pl}$ and plasmon-plasmon $\gamma_{pl,pl}$ emission rates. 
In the presence of extended media supporting conventional surface plasmon polaritons, $P_{a,nr}$ results in fluorescence quenching (decreased radiative far-field emission) of the emitter~\cite{anger2006}. Furthermore, both in this case and in  finite-size 2D nanostructures supporting localized surface plasmons, $P_{a,nr}$ also accounts for non-radiative mechanisms which could result in, {\it e.g.}, entangled lossy excitations %(double quenching)
\cite{rivera2016}. However, 
in low-dissipative systems such as the ones considered here, these excitations are negligible and plasmonic modes largely dominate the non-radiative emission channel. Provided that there are no resonant energy levels between $\hbar \omega_i$ and $\hbar \omega_f$, then $\gamma_0(\omega)$ and $t_{ab}(\omega)$ are broad-band and the TPSE channels inherit their main spectral characteristics directly from $P_{a,r}$ and $P_{a,nr}$. When the electromagnetic fields radiated by the quantum emitter and by the induced multipoles on the nanostructure are in (out of) phase, far-field constructive (destructive) interference occurs, rendering an asymmetric Fano-like profile for $P_{a,r}$. On the other hand, non-radiative processes are governed by absorption in the nanostructure, where the induced fields are much stronger than the emitter's. Hence, emitter-multipoles interferences are negligible and $P_{a,nr}$ results symmetric around resonances.
\begin{figure}
\centering
\includegraphics[width=1\linewidth]{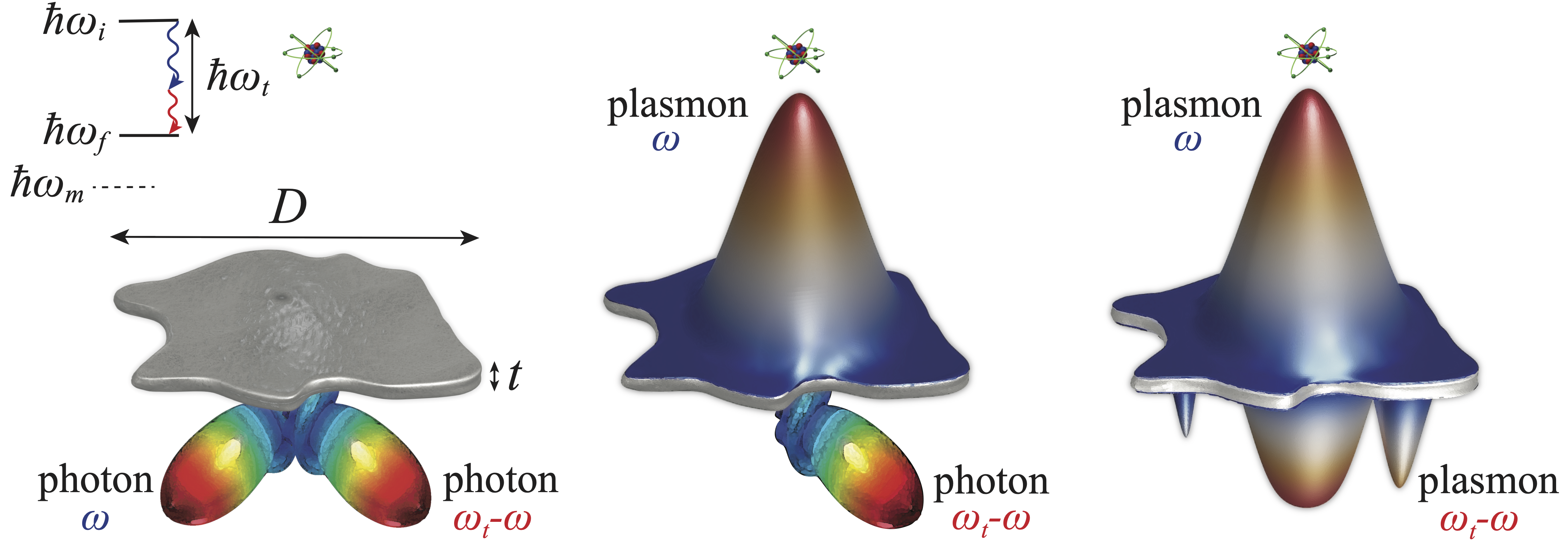}
\caption{Schematics of the system under study and representation of the TPSE pathways for a multi-level quantum emitter close to a 2D plasmonic nanostructure:  a pair of photons is emitted to the far-field (left), a hybrid photon-plasmon state is generated (center), or two plasmonic excitations are launched on the nanostructure (right). In each case the two-quanta states can be entangled in time-energy, linear, or angular momentum.}
\label{Fig1}
\end{figure}
\begin{figure}
\centering
\includegraphics[width=1\linewidth]{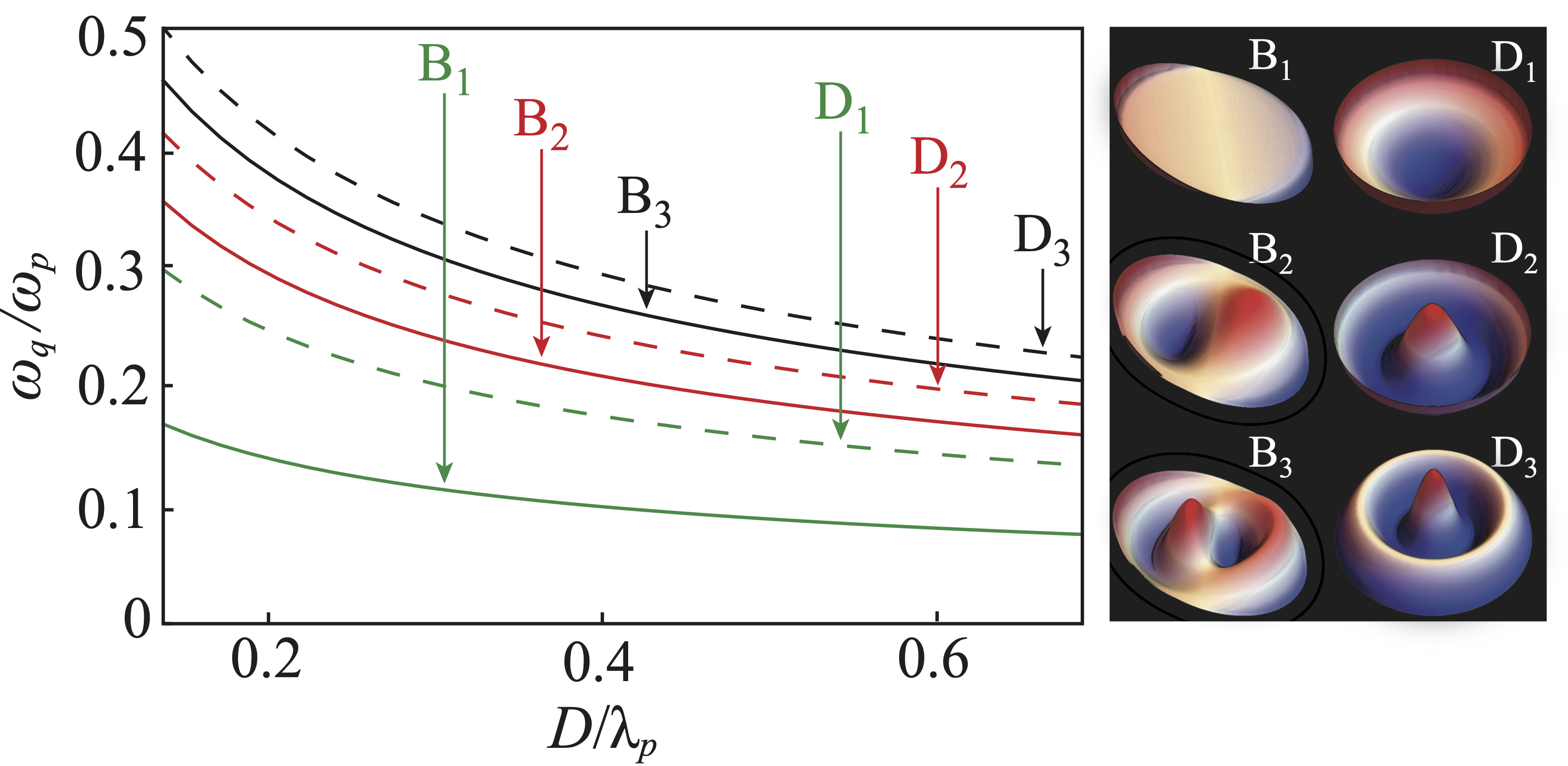}
\caption{Resonant frequencies $\omega_q$ for the three lowest energy bright ($B_q$) and dark ($D_q$) modes versus the diameter $D$  of a bilayer Ag(111) nano-disk ~\cite{maniyara2019, el-fattah2019}.  The corresponding spatial charge distributions are shown on the right panel. We model the optical response using a 2D Drude  conductivity $\sigma(\omega) = i\epsilon_0\omega_p^2t/(\omega+i/\tau)$, where $\hbar \omega_p =  2\pi  \hbar c/\lambda_p = 9.1$ eV and $\hbar \tau^{-1} = 18$ meV are the plasma frequency and relaxation rate of bulk Ag, and $t$ is the thickness of the nanostructure.}
\label{Fig2}
\end{figure}

To validate the above reasoning, we calculate the spectral lineshapes of the TPSE channels by employing a theoretical approach based on the plasmon wave function formalism~\cite{manjavacas2015, yu2017} (see supporting information~\cite{SI}). We consider 2D nanostructures supporting electromagnetic modes with resonant wavelengths much larger than their characteristic geometrical length scales (denoted as $D$), in which case it is sufficient to determine the electric field on the nanostructure in the quasi-static limit. In this regime, the nanostructure's plasmonic modes and  resonant frequencies (denoted as $\omega_q$) form an eigensystem that satisfies the Poisson equation. While the field modes and the corresponding charge density distribution depend only on the shape of the 2D structure, the resonant frequencies are affected by both the size $D$ and the conductivity $\sigma(\omega)$ of the material.  To compute the Purcell factors we use the identity \cite{novotny2012} $P_a(\omega) = W_a(\omega)/W_0(\omega) $,  where $W_a$ is the total power dissipated by a classical  electric dipole ${\bf d}_a = d\hat{{\bf e}}_a$, and $W_0$ is the corresponding dissipated power in free-space. 
 While the non-radiative part of $P_a$ can be computed through the total power absorbed by the nanostructure, the radiative component is mainly dominated by dipolar radiation~\cite{novotny2007}. A detailed derivation of the exact expressions for $P_{a,nr}$ and $P_{a,r}$ near arbitrary nanostructures can be found in Ref. \citenum{SI}.  When the conductivity of the metallic nanostructure is described by a low-loss Drude model, one can approximate the Purcell factors via a superposition of spectrally localized resonances. Hence,
\begin{equation}
P_{a,nr}(\omega) \simeq \sum_{q = 1}^{N}\frac{A_{a,q}}{\omega^2}\frac{(1 /2\tau)^2}{(\omega - \omega_q)^2 + (1/2\tau)^2}\, ,
\label{FinalPnr}
\end{equation}
which is a combination of Lorentzian line-shapes symmetric around each of the  $N$ distinct plasmonic resonances $\omega_q$   within the TPSE spectral range, and $\tau$ is the electronic relaxation time. The $\omega^{-2}$ factor is essential to  describe the TPSE spectrum  near $\omega = 0$ and $\omega = \omega_t$; nevertheless, far from these frequencies it is a good approximation to replace it by $\omega_q^{-2}$. Similarly, the radiative Purcell factor can be expressed as a combination of symmetric Lorentzian and asymmetric Fano~\cite{miroshnichenko2010, lukyanchuk2010, limonov2017} profiles,
\begin{equation}
P_{a,r}(\omega)\! \simeq\! \sum_{q = 1}^{N}\frac{B_{a,q}(1/2\tau)^2+(\omega - \omega_q + f_{a,q} /2\tau)^2}{(\omega - \omega_q)^2 + (1/2\tau)^2}  - (N-1).
\label{FinalPrad}
\end{equation}
 Here, $f_{a,q}({\bf R}_e) = \omega_p^2\tau t h_{a,q}({\bf R}_e)/D\omega_q$  is the Fano factor, where $\omega_p$ is the plasma frequency of the material, $t$ is the thickness of the nanostructure, and $ h_{a,q}({\bf R}_e)$ is a geometry-aware function. The coefficients $A_{a,q}$, $B_{a,q}$, and $f_{a,q}$ contain information about degeneracies of $\omega_q$, and the last term in Eq. \eqref{FinalPrad} arises from the non-zero overlap among the Fano resonances. The radiative Purcell factor peaks approximately at $\omega_q+(2\tau f_{a,q})^{-1}$ around which the Fano term overwhelms the Lorentzian one, but near the Fano dip at $\omega_q-f_{a,q}/2\tau$ the Lorentzian term becomes relevant preventing complete inhibition of photon emission. By tailoring $f_{a,q}$ through geometry or material properties it is possible to either enhance or suppress the generation of far-field radiation via $\gamma_{ph,ph}$ or $\gamma_{ph,pl}$.
\begin{figure}
\centering
\includegraphics[width=1\linewidth]{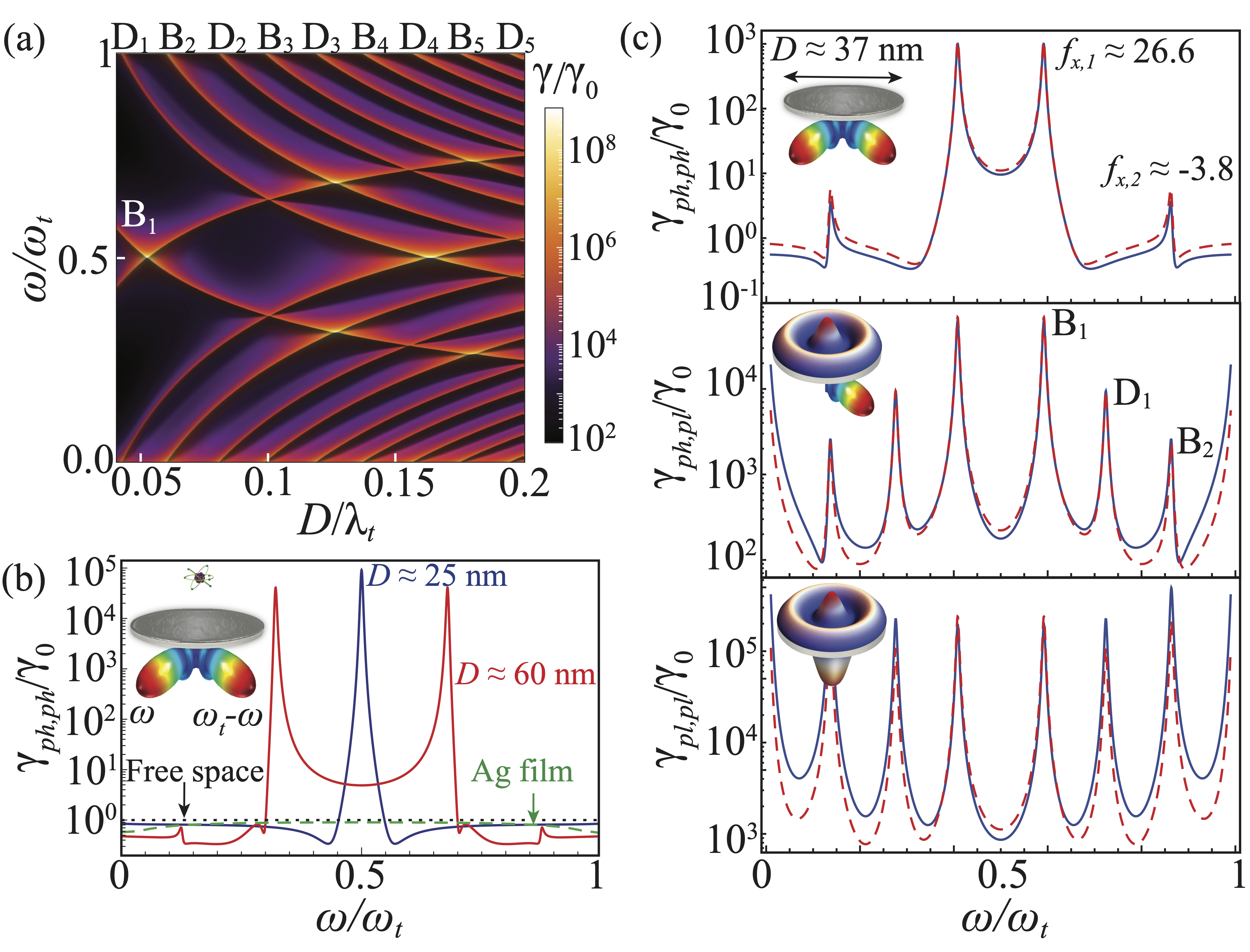}
\caption{(a) TPSE spectral density $\gamma(\omega)$ near a bilayer Ag nano-disk. The emitter is a quantum dot 
with transition frequency $\hbar \omega_t = 2.64$ eV 
placed at $z_e= 10$ nm.  (b) Photon-pair production rates for a Ag nano-disk (solid blue and red), a Ag film (green), and in free space (black). (c) TPSE spectral profiles for photon-photon (top), photon-plasmon (center), and plasmon-plasmon (bottom) decay channels. Solid (dashed) curves result from exact (approximated) calculations (see discussion in the text). The Fano asymmetry factor $f_{x,q}$ is displayed for the two bright resonances. }
\label{Fig3}
\end{figure}

We consider next a particular geometry amenable to analytical treatment, namely a plasmonic nano-disk~\cite{manjavacas2014} close to an on-axis quantum emitter. In this case the eigenmodes and eigenfrequencies supported by the nano-disk have a closed form~\cite{SI, fetter1986}, and   only azimuthally symmetric dark ($D_q$) and  dipolar bright  ($B_q$) modes can be excited.  The former ones do not radiate while the latter ones are able to 
leak into the far-field by emitting dipolar radiation~\cite{tamagnone2018}.  
Figure 2 depicts the resonant frequencies of the six lowest energy field modes versus the diameter of a metallic nano-disk, highlighting controlled optical response by properly choosing the structure's size.  The associated spatial charge distributions are also presented. The corresponding TPSE spectrum is shown in Fig. 3a, exhibiting a wealth of strongly localized peaks precisely along the curves for $\omega_q(D)$ and $\omega_t-\omega_q(D)$, and its maximum value is at $\omega_t/2$ ($2 \lambda_t \sim 940$ nm).  In our calculations, we consider spherically symmetric initial and final states~\cite{muniz2019,SI} in  Eq. (1) for which $t_{ab}(\omega) = \delta_{ab}/3$. In this case, the spectral enhancement lineshape of each emission channel   follow directly from the Purcell factors regardless of the emitter's intrinsic energy level structure. Single, dual, and even multi-band emissions are possible depending on the number of resonances below $\omega_t$. Cross-talk between bright-bright or dark-dark modes at complementary frequencies $\omega_{q'} (D)=\omega_t-\omega_q (D)$ produces extreme enhancements of the TPSE spectrum $\gamma(\omega)/\gamma_0(\omega) \sim 10^8$, while these are much smaller at dark-bright crossings. This results from the fact that, when the quantum emitter is on-axis, bright  and dark modes are effectively decoupled since they 
 can only be excited by virtual transition dipole moments parallel and orthogonal to the nanostructure, respectively. 
 
 Figure 3b compares $\gamma_{ph,ph}$ between confined and extended 2D metallic systems, evidencing that the finite size of the nanostructure is critical to accomplishing giant photon-photon production rates. 
Indeed, although a quantum emitter close to a metallic film experiences increased emission into surface-plasmon polaritons, these do not directly couple to photons, resulting in $ \gamma_{ph,ph}^{{\rm film}}/\gamma_0 \sim 1$. Contrarily, enhanced two-photon emission rates $\gamma_{ph,ph}^{{\rm disk}}/\gamma_{ph,ph}^{{\rm film}} \sim 10^ 5$ can be achieved in the plasmonic nanostructure since localized bright surface plasmons radiate into the far-field.  The spectral profiles of the TPSE channels are reported in Fig. 3c, where we observe a very good agreement between the TPSE lineshapes derived from the approximated expressions in  Eqs. (3), (4) and those obtained with full numerical evaluations of Eqs. (24), (28) of the supporting information. Close to plasmonic resonances there is a clear interplay of Fano and Lorentzian lineshapes that results in notable differences between the spectral profiles of $\gamma_{ph,ph}$
and those of the other emission mechanisms. The spectral distinction of $\gamma_{ph,pl}$ and $\gamma_{pl,pl}$ is more subtle: it is more prominent near the borders of the spectrum, and $f_{a,q}$ can be engineered to enhance their differences  around $\omega_t/2$.

In order to accomplish tunable TPSE rates we consider the nano-disk composed of active materials whose optical response can be dynamically controlled, {\it e.g.}, graphene~\cite{neto09}. Graphene not only provides the opportunity of emitting two-quanta in the mid-IR, but also allows for easier fabrication of 2D nanostructures as compared to metallic systems. Figure 4a reports $\gamma(\omega)$ for different Fermi energies of a graphene nano-disk, showing enhanced selective spectral emission (solid curves).  This is in stark contrast with the typical broadband spectrum achieved in monolayers~\cite{rivera2016} (dashed curves). Giant photon-pair production in this system is also possible, with  $\gamma_{ph,ph}^{{\rm disk}}/\gamma_{ph,ph}^{{\rm monolayer}} \gtrsim 10^9$ at the center of the spectrum (not shown). In Fig. 4b we address the question as to whether photon generation can be more efficient through two-photon transitions than via existing ordinary one-photon emission channels. The ratio between the probabilities of emitting at least one photon via two-photon transitions  and of generating a single photon via a one-photon process is presented in figure for the case of the nano-disk. These probabilities are computed through the TPSE quantum yield ${\rm QY}^{\rm TPSE}= (\gamma_{ph,ph}+\gamma_{ph,pl})/\gamma$ and the single photon quantum yield ${\rm QY}^{1q} = \gamma^{1q}_{ph}/\gamma^{1q}$, where $ \gamma^{1q}_{ph}$ is the radiative contribution to the one-quantum transition rate $\gamma^{1q}$. Serendipitously, we find that the fundamental dark mode $D_1$ acts as an amplifier of the TPSE photon-plasmon channel but as an attenuator of the photon one-quantum pathway. For frequencies near $\omega_{D_1}$, one-photon generation via TPSE is resonantly enhanced, being between two to four orders of magnitude larger than photon creation via  standard one-quantum emission.  On the other hand, there is also a broadband enhancement of ${\rm QY}^{\rm TPSE}/{\rm QY}^{1q}$ that takes place within regions of frequencies below $\omega_{B_1}$. These two kinds of enhancements are of a complete different nature. The resonant enhancement arises from the TPSE photon-plasmon emission channel that boosts ${\rm QY}^{\rm TPSE}$ via a non-radiative Lorentzian resonance, while ${\rm QY}^{1q}$ is spectrally flat and much smaller than ${\rm QY}^{\rm TPSE}$ near $\omega_{D_1}$.  The broadband enhancement results from the fact that ${\rm QY}^{1q}$ and ${\rm QY}^{\rm TPSE}$ have spectrally aligned resonant responses along the fundamental bright mode $B_1$, and as one moves to lower frequencies the former decreases faster than the latter.

Graphene nanostructures can also disrupt the usual unbalance between the total one- and two-quanta emission rates, making the latter competitive with the former through  tailoring the mobility $\mu$ or the Fermi energy $E_F$ of graphene.
\begin{figure}
\centering
\includegraphics[width=1\linewidth]{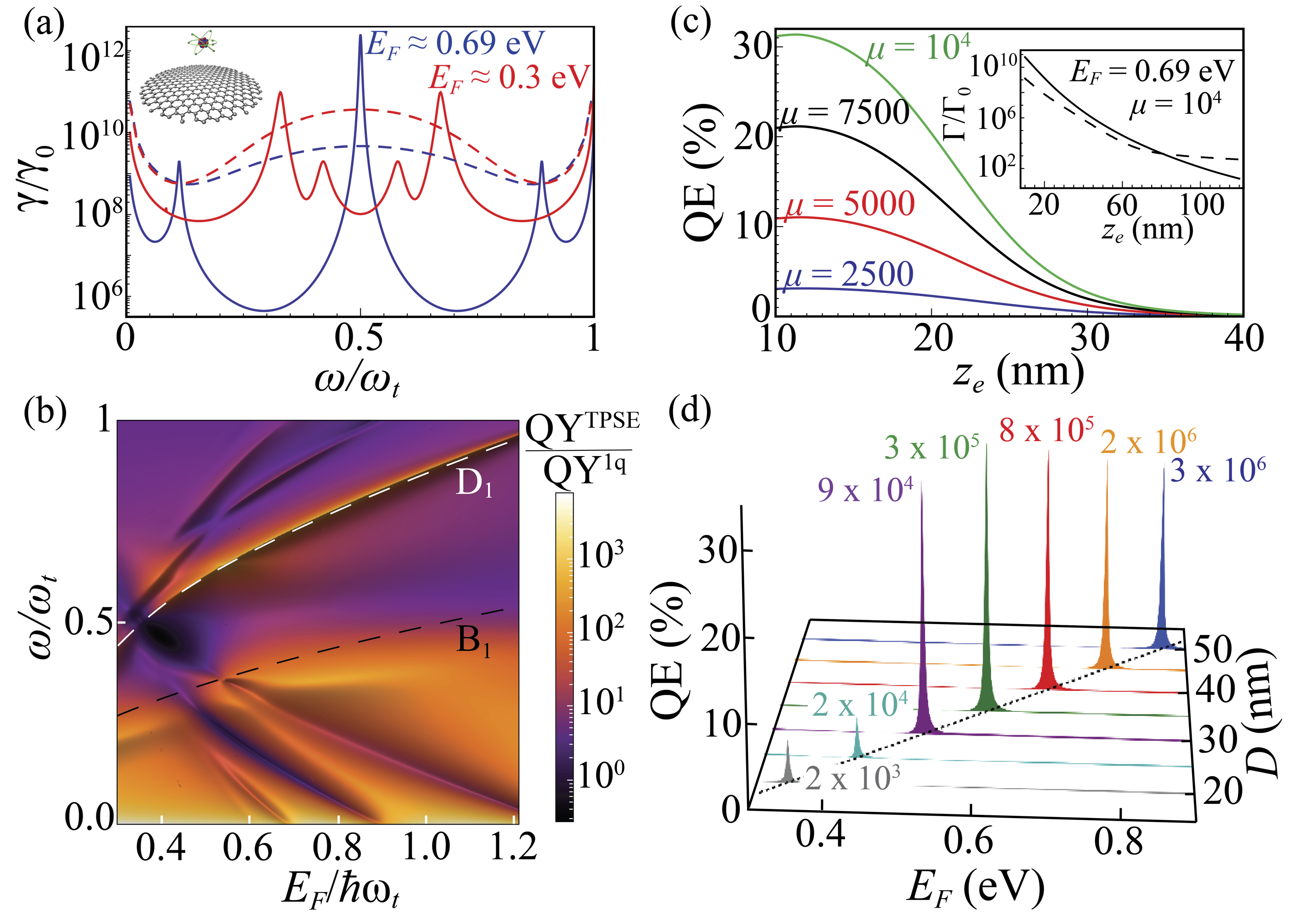}
\caption{ (a)  Spectral TPSE for a $D = 40$ nm graphene nano-disk (solid) and a graphene monolayer (dashed). The emitter ($\hbar \omega_t = 0.66$ eV)  is located at $z_e=10$ nm. Graphene's conductivity is modeled using intra- and inter-band contributions, mobility is $\mu=2500$ cm$^2$ V$^{-1}$ s$^{-1}$, and temperature is $T=300$ K.  (b)  Ratio of quantum yields between two- and one-quantum processes   for the nano-disk. (c)  Quantum efficiency versus distance for the TPSE $4s\rightarrow 3s$ transition in hydrogen ($\mu$ is in units of cm$^2$V$^{-1}$ s$^{-1}$). Inset: TPSE rate versus $z_e$ for the nano-disk (solid) and monolayer (dashed). (d) QE as a function of $E_F$ and $D$. The numbers near each QE profile show the photon-photon Purcell factor $\Gamma_{ph,ph}/\Gamma_0$, where 
$\mu=10^4$ cm$^2$ V$^{-1}$ s$^{-1}$ and $\Gamma_0$ is the free-space TPSE rate.}
\label{Fig4}
\end{figure}
For example, for a hydrogen emitter initially prepared in its $4s$ state,  $\Gamma_{4s \rightarrow 3s} \simeq 1.9 \times 10^8$ s$^{-1}$  while  the fastest competing one-quantum electric dipole transition gives $\gamma^ {1q}_{4s \rightarrow 3p} \simeq 1.2 \times 10^8$ s$^{-1}$ for a graphene nano-disk at a distance $z_e = 10$ nm from the emitter ($D = 40$ nm, $E_F = 0.69$ eV, and ultra-high mobility \cite{bolotin2008, dean2010} $\mu=10^4$ cm$^2$ V$^{-1}$ s$^{-1}$). In Fig. 4c,d we compare the TPSE rate for the $4s\rightarrow 3s$ transition in a hydrogen emitter with the competing one-quantum emission pathways. For  $z_e \lesssim 20$ nm the quantum efficiency  ${\rm QE} = \Gamma_{4s\rightarrow3s}/(\Gamma_{4s\rightarrow3s}+\gamma_{4s\rightarrow3p}^{1q}+\gamma_{4s\rightarrow2p}^{1q})$ 
reaches values $\sim 30\%$, which are much higher than in  graphene monolayers~\cite{rivera2016}.  Also, for distances $z_e \lesssim 80$ nm the total TPSE rate is larger in graphene nanostructures than in monolayers, highlighting that  the finite-size of the system is pivotal to achieving giant emission rates. 
For any disk diameter the QE can also be controlled by changing the Fermi energy, with optimized performance when $\omega_{B_1}(D) = \omega_t/2$, 
(dotted curve in the $(E_F,D)$ plane in Fig. 4d).  For $E_F\lesssim \hbar\omega_t/2$, interband transitions in graphene lead to the generation of entangled electron-hole pairs, which dominate over plasmonic excitations and suppress the total TPSE (left-most two peaks). 

 Experimental setups such as those of Refs. \citenum{ota2011,wang2019} can be employed to measure the TPSE at near-infrared frequencies from quantum dots with biexciton-exciton transitions. Recently developed synthesis techniques \cite{maniyara2019, el-fattah2019} can be employed to fabricate ultra-thin noble-metal nanostructures on a SiO$_2$-GaAs membrane with an embedded emitter layer.  For an InGaAs quantum dot located on-axis near a Ag nanodisk ($t = 1.65$ nm, $D = 62$ nm, SiO$_2$ thickness $ = 30$ nm), for example, the fundamental bright mode is excited at~\cite{note3} $\omega_{B_1} = \omega_t/2 \simeq 1.4$ eV, resulting in two-photon enhancements $\gtrsim10^4$. Such a giant enhancement 
is well above existing experimental sensitivities and should be easily detected (much smaller values $\sim10$ have already been measured in Ref. \citenum{wang2019}). High-resolution ($\sim 1\ \mu$eV) spectrometers can be used to scan the far-field TPSE spectral density and probe some of the predicted Fano and Lorentzian features. For instance, the Fano asymmetry factor can be obtained by reconstructing the two-photon spectrum via hyperspectral photon-coincidence measurements~\cite{hayat2008,wang2019} using near-infrared monochromators and photo-detectors.  Lorentzian signatures present in the hybrid photon-plasmon channel can be probed via frequency-resolved photoluminescence detection. Finally, these experiments combined with time-resolved fluorescence measurements \cite{lodhal2004, hoang2015} of the emitter's dynamics allow to extract the full TPSE rate $\Gamma$  in Eq \eqref{GammaTotal} and the individual decay probabilities for the three emission channels. 

  In summary, we have investigated 2D plasmonic nanostructures as a new platform for tailoring and enhancing TPSE. The strongly localized surface plasmons in these systems boost the TPSE beyond what is feasible in monolayers and 3D structures. The observation of the herein predicted TPSE effects is within experimental reach, and production rates of entangled photons much higher than those  achieved through parametric down conversion or spontaneous decay of bulk semiconductor emitters should be possible.  We developed a comprehensive theoretical toolbox to unravel the dominant emission channels, valid for finite-sized 2D systems with arbitrary geometric and material properties. We envision that our discovery of enhanced generation of photons via two-quanta decay in comparison to one-photon processes may lead to new nano-optics technologies. Altogether, our findings highlight the potential that TPSE in 2D plasmonic nanostructures has for photonics. This includes the active control of single-to-multiband emission spectra for sensing and spectroscopy functionalities, rapid generation of two-photon hyper-entangled states~\cite{gao2010} for quantum cryptography, as well as opportunities to develop novel infrared  two-quanta sources with high quantum efficiencies.

\begin{acknowledgments}
Research presented in this article was supported by the Laboratory Directed Research and Development program of Los Alamos National Laboratory under projects number 20190574ECR and 20180062DR. W.K.K. and Y.M. thank the Center for Nonlinear Studies at LANL for partial financial support. Y.M. and. C.F. acknowledge funding by the Coordena\c c\~ ao de Aperfei\c coamento de Pessoal de N\'ivel Superior (CAPES) and  Conselho Nacional de Desenvolvimento Cient\'ifico e Tecnol\'ogico (CNPq). A.M. acknowledges the National Science Foundation (grants ECCS-1710697 and DMR-1941680) and the UNM Center for Advanced Research Computing.

\textsuperscript{*} Correspondence: kortkamp@lanl.gov

\end{acknowledgments}

%%%%%%%%%%%%%%%%%%%%%%%%%%%%%%%%%%%%%%%%%%%%%%%%%%%%%%%%%%%%%%%%%%%%%
\let\oldsection\section

\onecolumngrid 
\appendix
\setcounter{equation}{0}

\pagebreak
\section{{\large Supporting Information \\ Two-photon spontaneous emission in atomically thin plasmonic nanostructures}}
 %\nopagebreak
%
\section{Two-photon spontaneous emission rate}

Here we present a short derivation of the two-photon spontaneous emission (TPSE) rate. The Hamiltonian of the system  is given by  $H = H_{A} + H_F + H_{int}$, where $H_{A}$ and $H_F$  are the emitter's and field's free Hamiltonians, respectively, and $H_{int}$ accounts for the emitter-field interaction. We assume that the dominant  transition wavelengths are much larger than the emitter dimensions, so that one can describe the emitter within  the electric dipole approximation. In this case ~\cite{milonni2013}
\begin{equation}
H_{int} = -{\bf d}\cdot{\bf E}({\bf R}_e) = -i \!\sum_\beta \!\!\sqrt{\frac{\hbar\omega_\beta}{2\epsilon_0}}\!\left[a_\beta {\bf d}\cdot{\bf A}_\beta({\bf R}_e) \! -\!  a^\dagger_\beta {\bf d}\cdot{\bf A}^\ast_\beta({\bf R}_e)\right]\! ,
\end{equation}
where ${\bf d}$ is the dipole moment operator, $a_{\beta}$ and  $a_{\beta}^{\dagger}$ are the field's annihilation and creation operators, ${\bf R}_e = {\bf r}_e + z_e\hat{\bf {z}}$ is the emitter's position, and $\{{\bf A}_\beta\}$ is a complete set of solutions of the Helmholtz equation subjected to appropriate boundary conditions. The TPSE rate can be calculated via second-order perturbation theory~\cite{sakurai2014,thiru1984}. By considering that initially the emitter is in an excited state  $|i\rangle $ and the field in the vacuum state, and that the final state corresponds to an emitter in a lower energy state  $|f\rangle $ with the field in a two-photon state, one obtains~\cite{Muniz2019}
\begin{equation}
\Gamma({\bf R}_e) = \frac{\pi}{4\epsilon_0^2\hbar^2} 
\sum_{\beta,\beta'}\omega_{\beta}\omega_{\beta'}|{\bf A}_{\beta}({\bf R}_e)\cdot \mathbb{D}(\omega_{\beta},\omega_{\beta'})
 \cdot {\bf A}_{\beta'}({\bf R}_e)|^2\delta(\omega_{\beta} + \omega_{\beta'} -\omega_t)\, .
\label{GammaModos}
\end{equation}
Here,  $\omega_t = ( E_f - E_i)/\hbar$ is the transition frequency between the initial and final states and we defined the tensor
\begin{equation}
\mathbb{D}(\omega_{\beta},\omega_{\beta'}) := \sum_m\left[\frac{{\bf d}_{im}{\bf d}_{mf}}{\omega_{im} - \omega_{\beta}} + \frac{{\bf d}_{mf}{\bf d}_{im}}{\omega_{im} - \omega_{\beta'}}\right], \label{diadico}
\end{equation}
with  ${\bf d}_{mm'} := \langle m|{\bf d}| m'\rangle$, $\omega_{mm'} := (E_m - E_{m'})/\hbar$, and the summation extends over all the emitter's intermediate states ($m \neq i, f$). We can also conveniently express the TPSE rate in Eq. (\ref{GammaModos}) in terms of the Green's tensor $\mathbb{G}({\bf R},{\bf R}^\prime;\omega)$ of the Helmholtz's equation by using its spectral representation~\cite{novotny2012}, 
\begin{equation}
\mathrm{Im}\mathbb{G}({\bf R}, {\bf R}^\prime;\omega) = \frac{\pi c^2}{2\omega}\sum_{\beta}{\bf A}_{\beta}({\bf R}){\bf A}_{\beta}^\ast({\bf R}^\prime)\delta(\omega - \omega_\beta),
\end{equation}  
which leads to
\begin{eqnarray}
\Gamma({\bf R}_e) = \int_0^{\omega_t} d\omega\frac{\mu_0^2}{\pi\hbar^2}\omega^2(\omega_t - \omega)^2\mathrm{Im}\mathbb{G}_{il}({\bf R}_e,{\bf R}_e;\omega)\mathrm{Im}\mathbb{G}_{jk}({\bf R}_e,{\bf R}_e;\omega_t - \omega)\mathbb{D}_{ij}(\omega,\omega_t - \omega)\mathbb{D}_{lk}^\ast(\omega,\omega_t - \omega). \label{GammaGreen}
\end{eqnarray}
This expression for the TPSE rate is valid regardless of the base of choice to express the Green function.

We will now relate the TPSE spectral density $\gamma({\bf R}_e,\omega)$(the integrand of Eq. \eqref{GammaGreen}) to the local density of states (LDOS). This is possible by noting that 
$\mathrm{Im}\mathbb{G}({\bf R}_e,{\bf R}_e; \omega)$ is a real and symmetric matrix~\cite{buhmann2012}, therefore it can be diagonalized. For systems where the basis $\{{\bf \hat{e}}_a\}$ ($a = 1, \ 2,\ 3$) which diagonalizes $\mathrm{Im}\mathbb{G}({\bf R}_e,{\bf R}_e;\omega)$ is the same at complementary frequencies $\omega$ and $\omega_t-\omega$,
we have $\mathrm{Im}\mathbb{G}_{ab}({\bf R}_e, {\bf R}_e;\omega) = \mathrm{Im}\mathbb{G}_{aa}({\bf R}_e, {\bf R}_e;\omega)\delta_{ab}$, therefore
\begin{equation}
\gamma({\bf R}_e,\omega) = 
\left(\frac{6\pi c}{\omega}\right)^2\gamma_0(\omega)\sum_{a,b}t_{ab}(\omega)
\mathrm{Im}\mathbb{G}_{aa}({\bf R}_e, {\bf R}_e;\omega)\mathrm{Im}\mathbb{G}_{bb}({\bf R}_e, {\bf R}_e;\omega), 
\end{equation}
where we defined $t_{ab}(\omega) = |\mathbb{D}_{ab}(\omega,\omega_t - \omega)|^2/|\mathbb{D}(\omega,\omega_t - \omega)|^2$ 
with $\vert\mathbb{D}(\omega,\omega_{0} - \omega)\vert^2 = \mathbb{D}_{ab}(\omega,\omega_{0} - \omega)\mathbb{D}^\ast_{ab}(\omega,\omega_{0} - \omega)$. Also, $\gamma_0(\omega) = \mu_0^2\omega^3(\omega_{0} - \omega)^3\vert\mathbb{D}(\omega,\omega_{0} - \omega)\vert^2/36\pi^3\hbar^2c^2$ is the free-space spectral density. 
 Once we recall the relation between the Purcell factor for a dipole moment oriented along the ${\bf \hat{e}}_a$-direction, namely~\cite{novotny2012}
\begin{equation}
P_a({\bf R}_e,\omega) = \frac{6\pi c}{\omega} \mathrm{Im}\mathbb{G}_{aa}({\bf R}_e, {\bf R}_e;\omega),
\label{PurcellFactors}
\end{equation}
%
%{\color{red}\sout{as the  Purcell factor for a dipole moment oriented along the ${\bf \hat{e}}_a$-direction~\cite{novotny2012}.}}
we obtain
\begin{equation}
\gamma({\bf R}_e,\omega) = 
\gamma_0(\omega)\sum_{a,b}t_{ab}(\omega)
P_a({\bf R}_e,\omega)P_b({\bf R}_e,\omega_t-\omega). 
\label{gamma/gamma_0}
\end{equation}
Equations (\ref{gamma/gamma_0}) and (\ref{PurcellFactors}) establish an explicit relation between the TPSE and the local density of photonic states, which is proportional to the Purcell factor.

\section{Eigenmode expansion and Plasmon Wave Function Formalism}

We follow the plasmon wave function (PWF) formalism presented in Refs.~\citenum{abajo2015,yu2017} in order to obtain the charge distribution induced on a ultrathin metallic nanostructure due to an external electric field ${\bf E}^{ext}({\bf R}, \omega)$. This approach assumes a large mismatch between the characteristic size ($D$) of the metallic nanostructures and their resonant wavelengths ($\lambda_\alpha$), in which case the optical response of the system can be described in the electrostatic regime. In this limit, the parallel component of the electric field over the surface of the nanostructure satisfies
\begin{equation}
{\bf E}_\parallel({\bf r}, \omega) = {\bf E}^{ext}_\parallel({\bf r}, \omega) + \frac{i\sigma(\omega)}{4\pi\epsilon_0\omega}\nabla_{{\bf r}}\int \frac{d^2{\bf r}^\prime}{|{\bf r} - {\bf r}^\prime|}\nabla_{{\bf r}^\prime}\cdot f({\bf r}^\prime){\bf E}_\parallel({\bf r}^\prime, \omega).
\end{equation}
Here, $\sigma(\omega)$ is the surface conductivity of the nanostructure and $f({\bf r})$ is a filling function which is equal to $1$ when the in-plane 2D position vector ${\bf r}$ lies within the nanostructure and $0$ elsewhere. It is convenient to re-write the above equation in terms of the dimensionless variable ${\bf u} = {\bf r} /D$ and of  $\bm{\mathcal{E}}({\bf u}, \omega) = D\sqrt{f(D{\bf u})}{\bf E}_\parallel(D{\bf u}, \omega)$, namely
\begin{equation}
\bm{\mathcal{E}}({\bf u}, \omega) = \bm{\mathcal{E}}^{ext}({\bf u}, \omega) + \eta(\omega)\int d^2{\bf u}^\prime \,\mathbb{M}({\bf u},{\bf u}^\prime)\cdot\bm{\mathcal{E}}({\bf u}^\prime, \omega),  
\label{SelfConsistent}
\end{equation}
where 
\begin{equation}
\eta(\omega) = \frac{i\sigma(\omega)}{4\pi\epsilon_0\omega D} \,\,\, \mbox{and} \,\,\, \mathbb{M}({\bf u},{\bf u}^\prime) = \sqrt{f({\bf u})f({\bf u}^\prime)}\nabla_{{\bf u}}\nabla_{{\bf u}}\frac{1}{|{\bf u} - {\bf u}^\prime|}.
\end{equation}
$ \mathbb{M}({\bf u},{\bf u}^\prime) $ is a real and symmetric operator which depends only on the geometry of the nanostructure. Therefore, $ \mathbb{M}$ admits a complete set of eigenmodes ${\bf V}_\alpha({\bf u}) $ and real eigenvalues $1/\eta_\alpha$ defined by 
\begin{equation}
\int d^2{\bf u}^\prime \,\mathbb{M}({\bf u},{\bf u}^\prime)\cdot{\bf V}_\alpha({\bf u}^\prime) = \frac{1}{\eta_\alpha}{\bf V}_\alpha({\bf u})\, .
\label{EigenvalueEquation}
\end{equation}
The eigenmodes are the solutions of Eq. \eqref{SelfConsistent} in the absence of an external electromagnetic field, and give the electric field profile over the surface of the nanostructure. They also satisfy the following closure and orthogonality relations, respectively 
\begin{align}
\sum_\alpha {\bf V}^\ast_\alpha({\bf u})\otimes{\bf V}_\alpha({\bf u}^\prime) = \delta({\bf u} - {\bf u}^\prime)\mathbb{I}_2  \hspace{0.3cm} \textrm{and} \hspace{0.3cm} 
\int d^2{\bf u} \,{\bf V}^\ast_\alpha({\bf u})\cdot{\bf V}_{\alpha^\prime}({\bf u}) = \delta_{\alpha\alpha^\prime}.
\label{EigenmodesOrthogonality}
\end{align}

By expanding $\bm{\mathcal{E}}$ and $\bm{\mathcal{E}}^{ext}$ in terms of ${\bf V}_\alpha({\bf u}) $ and using Eq. \eqref{SelfConsistent} one obtains
\begin{equation}
\bm{\mathcal{E}}({\bf u}, \omega) = \sum_\alpha \frac{c_\alpha}{1 - \eta(\omega)/\eta_\alpha}{\bf V}_\alpha({\bf u}), \label{FieldOverSurface}
\end{equation}
where
\begin{equation}
c_\alpha = \int d^2{\bf u}\,{\bf V}^\ast_\alpha({\bf u})\cdot\bm{\mathcal{E}}^{ext}({\bf u}, \omega) \label{cj}\, .
\end{equation}
Eq. \eqref{FieldOverSurface} establishes that for any external field $\bm{\mathcal{E}}^{ext}$ the electric field over the nanostructure is a superposition of the eigenmodes ${\bf V}_\alpha$. Each eigenmode can be excited if the frequency of the external field matches one of the resonance frequencies of the system, given by $\mathrm{Re}[\eta(\omega_\alpha)]= \eta_\alpha$. Once one knows the field over the surface of the nanostructure, one can use Ohm's law, ${\bf K}({\bf r}, \omega) = \sigma(\omega)f({\bf r}){\bf E}_\parallel({\bf r}, \omega)$, together with the continuity equation, $i\omega\rho_{2D}({\bf r}, \omega) = \nabla_{{\bf r}}\cdot {\bf K}({\bf r}, \omega)$, to derive an expression for the charge density distribution. Hence, 
\begin{equation}
\rho_{2D}({\bf r}, \omega) = \frac{4\pi\epsilon_0}{D}\sum_\alpha \frac{c_\alpha}{1/\eta_\alpha - 1/\eta(\omega)}v_\alpha({\bf u}),
 \label{ChargeDensity}
\end{equation}
where we defined the plasmon wavefunction $v_\alpha({\bf u}) = \nabla_{\bf u}\cdot\sqrt{f({\bf u})}{\bf V}_\alpha({\bf u})$, which corresponds to the normalized charge distribution of the plasmon mode $\alpha$.
By taking the divergence of $\sqrt{f({\bf u})}$ times Eq. \eqref{EigenvalueEquation} one can show that the plasmon wave functions satisfy the Poisson equation:
\begin{equation}
 \nabla^2_{\bf u}\int d^2{\bf u}^\prime \dfrac{v_\alpha({\bf u}^\prime)} {|{\bf u} - {\bf u}^\prime|}  = \eta_\alpha^{-1} v_\alpha({\bf u}).
 \label{EigenvaluePWF}
 \end{equation}
Also, it follows from 
the previous equation that ${\bf V}_\alpha$ can be cast in terms of the corresponding PWF as
\begin{equation}
{\bf V}_\alpha({\bf u}) = \sqrt{f({\bf u})}\eta_\alpha\int d^2{\bf u}'\frac{v_\alpha({\bf u}')({\bf u} - {\bf u}')}{|{\bf u} - {\bf u}'|^3},
\end{equation}
and by taking into consideration the orthogonality condition for the eigenmodes, given in Eq. \eqref{EigenmodesOrthogonality}, one can then show that the PWFs must obey the following relation:
\begin{equation}
\int d^2{\bf u}\int d^2{\bf u}^\prime\frac{v_\alpha({\bf u})v_{\alpha^\prime}({\bf u}^\prime)}{|{\bf u} - {\bf u}^\prime|} = -\frac{\delta_{\alpha\alpha^\prime}}{\eta_\alpha}.\label{PWForthogonality}
\end{equation}
\section{Purcell factors due to a metallic nanostructure}

The Purcell factor 
$P_a({\bf R}_e,\omega)$ can be calculated with the aid of the identity \cite{novotny2012} $P_a({\bf R}_e,\omega) = W_a({\bf R}_e,\omega)/W_0(\omega) $,  where $W_a({\bf R}_e,\omega)$ is the total power dissipated by a classical dipole ${\bf d}_a = d_a{\bf \hat{e}}_a$ oscillating with frequency $\omega$ at position  ${\bf R}_e$ near the nanostructure, and $W_0(\omega)$ is the corresponding dissipated power in free space. As a consequence, one can write~\cite{novotny2012}
\begin{equation}
P_a({\bf R}_e,\omega) = P_{a,nr}({\bf R}_e,\omega) + P_{a,r}({\bf R}_e,\omega)\, ,
\label{DecayChannels}
\end{equation}
where
\begin{eqnarray}
P_{a,nr}({\bf R}_e,\omega) &=& \frac{6\pi\epsilon_0c^3}{\omega^4|{\bf d}_a|^2}\!\!\int d^3{\bf R}^\prime \mathrm{Re}\{{\bf J}^\ast({\bf R}^\prime,\omega)\cdot{\bf E}({\bf R}^\prime,\omega)\} \, ,\cr
P_{a,r}({\bf R}_e,\omega) &=& \frac{6\pi\epsilon_0c^3}{\omega^4|{\bf d}_a|^2}\!\!\int_{R^\prime \rightarrow \infty}\!\!\!\!\!\!\!\!\!\! d{\bf A}^\prime\cdot \mathrm{Re}\{{\bf E}({\bf R}^\prime,\omega)\times{\bf H}^\ast({\bf R}^\prime,\omega)\}
\label{DecayChannels2}
\end{eqnarray}
correspond to the contribution of  non-radiative  and radiative decay channels to the Purcell factor (LDOS), respectively.
    
We start by calculating the non-radiative contribution. By inserting ${\bf J}({\bf R}^\prime,\omega) = {\bf K}({\bf r}^\prime, \omega)\delta(z^\prime) = \sigma(\omega)f({\bf r}^\prime){\bf E}_\parallel({\bf r}^\prime, \omega)\delta(z^\prime)$ into equation \eqref{DecayChannels2} and using the orthogonality relation in Eq. \eqref{EigenmodesOrthogonality}, we derive
\begin{equation}
P_{a,nr}({\bf R}_e,\omega) = \frac{6\pi\epsilon_0c^3}{\omega^4|{\bf d}_a|^2} \mathrm{Re}[\sigma(\omega)]\sum_\alpha\bigg\vert \frac{c_\alpha}{1 - \eta(\omega)/\eta_\alpha} \bigg\vert^2 . 
\label{Step1}
\end{equation}
%
%{\color{red}\sout{By using that}} 
The external field over the nanostructure, which is given by the electric field generated by the dipole, can be approximated by  ${\bf E}^{ext}({\bf R}^\prime,\omega) = \frac{1}{4\pi\epsilon_0}\nabla {\bf d}_a\cdot \nabla|{\bf R}_e - {\bf R}^\prime|^{-1}$  in the near-field regime. By using this expression  in eq. \eqref{cj}, we obtain $c_\alpha = {\bf d}_a\cdot{\bf F}_\alpha^\ast({\bf R}_e)/4\pi\epsilon_0D^2$, where
\begin{equation}
{\bf F}_\alpha({\bf R}_e) = \int d^2{\bf u}'\frac{v_\alpha({\bf u}')({\bf R}_e/D - {\bf u}')}{|{\bf R}_e/D - {\bf u}'|^3} \label{FieldPWF}
\end{equation}
corresponds to the field generated at position ${\bf R}_e$ by the $\alpha$-th PWF mode. This result allows us to write Eq. (\ref{Step1}) as
\begin{equation}
P_{a,nr}({\bf R}_e,\omega) = \frac{3c^3}{2D^3\omega^3}\mathrm{Im}\sum_\alpha {\bf \hat{e}}_a\cdot \frac{{\bf F}_\alpha({\bf R}_e)\otimes{\bf F}^\ast_\alpha({\bf R}_e)}{1/\eta(\omega)-1/\eta_\alpha}\cdot{\bf \hat{e}}_a.
\label{PnrPWF}
\end{equation}
In the regime $D\ll \lambda_\alpha$, the radiative contribution to the Purcell factor is due to the system's emitted dipolar radiation, which can be well approximated by~\cite{novotny2007}
\begin{equation}
P_{a,r}({\bf R}_e,\omega)\simeq \frac{|{\bf d}_a + {\bf d}_{a,ind}({\bf R}_e,\omega)|^2}{|{\bf d}_a|^2}, 
\label{PrPWF}
\end{equation}
where 
\begin{equation}
{\bf d}_{a,ind}({\bf R}_e,\omega) = \int d^2{\bf r} \,{\bf r} \rho_{2D}({\bf r},\omega) = \sum_\alpha \frac{\bm{\zeta}_\alpha\otimes{\bf F}_\alpha^\ast({\bf R}_e)}{1/\eta(\omega)-1/\eta_\alpha}\cdot{\bf d}_a 
\label{InducedDipoleDipole}
\end{equation}
is the dipole moment induced in the nanostructure by the field of the dipole ${\bf d}_a$, and
\begin{equation}
\bm{\zeta}_\alpha = \int d^2{\bf u}\,{\bf u}v_\alpha({\bf u}) 
\label{DipolePWF}
\end{equation}
corresponds to the dipole moment of the plasmon $\alpha$. Therefore,
\begin{equation}
P_{a,r}({\bf R}_e,\omega) = \bigg\vert {\bf \hat{e}}_a + \sum_\alpha \frac{\bm{\zeta}_\alpha\otimes{\bf F}_\alpha^\ast({\bf R}_e)}{1/\eta(\omega)-1/\eta_\alpha }\cdot{\bf \hat{e}}_a \bigg\vert^2 .
\label{Prad2}
\end{equation}
It is important to note that Eqs. \eqref{PnrPWF} and \eqref{Prad2} are exact (within the dipole approximation for the nanostructure) expressions for the non-radiative and radiative Purcell factors, and can be numerically evaluated for any material once the PWFs for a given geometry are known.
\section{Spectral line-shape of two-photon decay channels}

Following the main text, we assume that the excited and ground states can be well described by  $s-$ orbital wavefunctions, in which case, Eq. (\ref{gamma/gamma_0}) simplifies to $\gamma({\bf R}_e,\omega)/\gamma_0(\omega) = \sum_a P_a({\bf R}_e,\omega)P_a({\bf R}_e,\omega_t-\omega)/3$. Note that this quantity is agnostic with respect to the emitter's electronic structure, and it depends on the emitter properties only through the transition frequency $\omega_t$. By taking advantage of Eq. (\ref{DecayChannels}), we can identify the spectral enhancements associated to the plasmon-plasmon, photon-plasmon, and photon-photon decay channels of the 
two-quanta decay process, namely
\begin{align}
\frac{\gamma_{pl,pl}({\bf R}_e,\omega)}{\gamma_0(\omega)} &= \frac{1}{3}\sum_aP_{a,nr}({\bf R}_e,\omega)P_{a,nr}({\bf R}_e,\omega_t-\omega), \label{Pnrnr}
\\
\frac{\gamma_{ph,pl}({\bf R}_e,\omega)}{\gamma_0(\omega)} &= \frac{1}{3}\sum_a\left[P_{a,nr}({\bf R}_e,\omega)P_{a,r}({\bf R}_e,\omega_t-\omega) + P_{a,r}({\bf R}_e,\omega)P_{a,nr}({\bf R}_e,\omega_t-\omega)\right], \label{Prnr}
\\
\frac{\gamma_{ph,ph}({\bf R}_e,\omega)}{\gamma_0(\omega)} &= 
\frac{1}{3}\sum_a
P_{a,r}({\bf R}_e,\omega)P_{a,r}({\bf R}_e,\omega_t-\omega). \label{Prr}
\end{align}

In order to investigate the line-shape of these spectral enhancements, we consider that the conductivity of the nanostructure is well described by a Drude model  $\sigma(\omega) = i\epsilon_0\omega_p^2t/(\omega + i/\tau)$, where $\omega_p$ is the bulk plasma frequency, $t$ is the thickness of the nanostrucutre, and $\tau$ ($\gg 1/\omega_p$) accounts for dissipation in the system. The resonant frequencies $\omega_\alpha$ of the nanostructure are determined through $\textrm{Re}[1/\eta_\alpha - 1/\eta(\omega_\alpha)] =0$, resulting in $ \omega_\alpha \simeq \sqrt{-\omega_p^2t/4\pi D \eta_\alpha}$ in our problem. Using this expression in Eq. (\ref{PnrPWF}) we find that the non-radiative Purcell factor can be written as
\begin{equation}
P_{a,nr}({\bf R}_e,\omega) =  \sum_{q = 1}^{N}\frac{3c^3\omega_p^2t}{8\pi D^4 \omega^2\tau} \frac{\sum_{j = 1}^{g_q}|{\bf \hat{e}}_a\cdot{\bf F}_{q,j}({\bf R}_e)|^2}{(\omega^2 - \omega_q^2)^2 + \omega^2/\tau^2}\, ,
\label{Pnrdegenerate}
\end{equation}
where we have split the summation over modes $\alpha$ into a sum in $q$ over all the $N$ resonances present in the TPSE spectrum, and a sum in $j$ over the degenerate modes. In Eq. (\ref{Pnrdegenerate}) $g_q$ is the degree of degeneracy of the $q$-th resonance.

In the regime of small dissipation, the overlap between different resonances is negligible, and we can expand each term in the above sum around the corresponding $\omega_q$, leading to  
\begin{equation}
P_{a,nr}({\bf R}_e,\omega) \simeq \sum_{q = 1}^{N}\frac{A_{a,q}}{\omega^2}\frac{(1/2\tau)^2}{(\omega - \omega_q)^2 + (1/2\tau)^2}\, , 
\label{FinalPnr}
\end{equation}
where
\begin{equation}
A_{a,q} = \frac{3c^3\omega_p^2t\tau}{8\pi D^4\omega_q^2} \sum_{j = 1}^{g_q}|{\bf \hat{e}}_a\cdot{\bf F}_{q,j}({\bf R}_e)|^2\, .
\end{equation}
We kept the prefactor $1/\omega^2$ since it comes from the normalization by the free space spectral density and determines the spectrum behaviour near $\omega = 0$ and $\omega = \omega_t$. It should be noticed, however, that far from $\omega = 0$ (and $\omega = \omega_t$) it is a good approximation to replace $1/\omega^2$ by $1/\omega_q^2$. Note that any eigenmode supported by the system provides a Lorentzian line-shape for the non-radiative part of the spectrum regardless of the geometry of the nanostructure. Also, precisely at a given plasmon resonance $\omega_{q^\prime}$, the non-radiative Purcell factor reduces to 
$ P_{a,nr}({\bf R}_e,\omega_{q^\prime}) = (6\pi c^3\eta^2_{q^\prime}\tau/{D^2 \omega_p^2 t})\sum_{j = 1}^{g_q}|{\bf \hat{e}}_a\cdot{\bf F}_{q^\prime,j}({\bf R}_e)|^2 $. We comment that the particular case of graphene can be obtained by replacing $\omega_p^2t \rightarrow e^2E_F/\pi\epsilon_0\hbar^2$ and $\tau \rightarrow E_F\mu/ev_F^2$, where $v_F$ is the Fermi velocity, $E_F$ is the Fermi energy, and $\mu$ is the charge carrier's mobility. This gives a non-radiative contribution at resonance proportional to $\mu/D^2$, being independent of $E_F$. 

Now we will do the same analysis for the radiative contribution given by Eq. \eqref{Prad2}. To write $P_{a,r}$ as a sum over resonances is more subtle than the previous case since $\displaystyle{\lim_{\omega \to \infty }P_{a,r} \rightarrow 1}$, which means that there is always an overlap between different resonances due to the free space contribution. Therefore, in order to write $P_{a,r}$ as a sum of functions which accurately describe each resonance near its own resonance frequency, we must subtract the contribution from all other $N - 1$ resonant terms. Hence,
\begin{equation}
P_{a,r}({\bf R}_e,\omega) \simeq \sum_{q = 1}^{N}\bigg\vert {\bf \hat{e}}_a + \frac{\omega_p^2t}{4\pi D}\frac{\sum_{j = 1}^{g_q}{\bf \hat{e}}_a\cdot {\bf F}_{q,j}^\ast({\bf R}_e)\otimes \bm{\zeta}_{q,j}}{\omega^2 - \omega_q^2 + i\omega/\tau}\bigg\vert^2 - (N - 1).
\end{equation}
Expanding the denominator of each resonant term around its corresponding $\omega_q$ yields
\begin{equation}
P_{a,r}({\bf R}_e,\omega) = \sum_{q = 1}^{N}\frac{\bigg\vert (\omega - \omega_q + i/2\tau){\bf \hat{e}}_a + \frac{\omega_p^2t}{8\pi D\omega_q}\sum_{j = 1}^{g_q}{\bf \hat{e}}_a\cdot {\bf F}_{q,j}^\ast({\bf R}_e)\bm{\zeta}_{q,j}\bigg\vert^2}{(\omega - \omega_q)^2 + (1/2\tau)^2} - (N - 1).
\end{equation}
Finally, we express $\bm{\zeta}_{q,j}$ in terms of its components parallel and perpendicular to the dipole moment, $\bm{\zeta}_{q,j} = \bm{\zeta}_{a;q,j}^{\parallel} + \bm{\zeta}_{a;q,j}^\perp$, where $\bm{\zeta}_{a;q,j}^\parallel = (\bm{\zeta}_{q,j}\cdot{\bf \hat{e}}_a){\bf \hat{e}}_a$ and $\bm{\zeta}_{a;q,j}^\perp = \bm{\zeta}_{q,j} - (\bm{\zeta}_{q,j}\cdot{\bf \hat{e}}_a){\bf \hat{e}}_a$. This results in 
\begin{equation}
P_{a,r}({\bf R}_e,\omega) = 1+\sum_{q = 1}^{N}\frac{(\omega - \omega_q + f_{a,q}/2\tau)^2+B_{a,q}\times (1/2\tau)^2}{(\omega - \omega_q)^2 + (1/2\tau)^2}  - N ,
\label{FinalPrad}
\end{equation}
where
\begin{equation}
f_{a,q} = \frac{\omega_p^2\tau t}{4\pi D\omega_q}\sum_{j = 1}^{g_q} \mathrm{Re}\left[{\bf \hat{e}}_a\cdot{\bf F}_{q,j}^\ast({\bf R}_e)\zeta_{a;q,j}^\parallel\right]
\end{equation}
is the Fano asymmetry parameter of the $q$-th radiative resonance, and
\begin{equation}
B_{a,q} = \left[ 1 + \frac{\omega_p^2t}{8\pi D\omega_q}\sum_{j = 1}^{g_q} \mathrm{Im}\left[{\bf \hat{e}}_a\cdot{\bf F}_{q,j}^\ast({\bf R}_e)\zeta_{a;q,j}^\parallel\right]\right]^2 + \bigg\vert\frac{\omega_p^2t}{8\pi D\omega_q}\sum_{j = 1}^{g_q} {\bf \hat{e}}_a\cdot{\bf F}_{q,j}^\ast({\bf R}_e)\zeta_{a;q,j}^\perp\bigg\vert^2
\end{equation}
is the amplitude of the Lorentzian resonance. Therefore, by using Eqs. (\ref{FinalPnr}) and (\ref{FinalPrad}) in Eqs (\ref{Pnrnr})-(\ref{Prr}) we are able to fully describe the line-shape of the spectral enhancements for each decay channel in the TPSE process.

\section{Plasmon wave functions modes for a nanodisk}

In this section we obtain the PWFs directly from equation \eqref{EigenvaluePWF} for the case of a nanodisk. The PWFs in polar coordinates can be written as a radial function $R_{ln}(u)$ times an angular function of the form $e^{il\phi}$. We also expand the term $|{\bf u} - {\bf u}^\prime|^{-1}$ in terms of Bessel functions, namely~\cite{fetter1986}
\begin{equation}
\frac{1}{|{\bf u} - {\bf u}^\prime|} = \int_0^\infty dp \sum_{m = -\infty}^{\infty}J_{|m|}(up)J_{|m|}(u^\prime p)e^{im(\phi - \phi^\prime)}. \label{Step3}
\end{equation}
In this way, equation \eqref{EigenvaluePWF} reduces to
\begin{equation}
2\pi\nabla^2_{\bf u}\int_0^{1/2} du^\prime u^\prime \int_0^\infty dp J_{|l|}(up)J_{|l|}(u^\prime p)R_{ln}(u^\prime)e^{il\phi} = \frac{1}{\eta_{ln}} R_{ln}(u)e^{il\phi}.
\end{equation}
In order to deal with the Laplacian, we write the right-hand side of the previous equation as $\frac{1}{\eta_{ln}} \int_0^{1/2}du^\prime u^\prime R_{ln}(u^\prime) \frac{\delta(u^\prime - u)}{u^\prime}e^{il\phi}$ and then recall that the Green's function of the radial part of the Poisson equation in cylindrical coordinates satisfies $\nabla^2_{\bf u}G_l(u,u^\prime)e^{il\phi} = -\frac{\delta(u^\prime - u)}{u^\prime}e^{il\phi}$. Hence,
\begin{equation}
\int_0^{1/2} du^\prime u^\prime \int_0^\infty dp J_{|l|}(up)J_{|l|}(u^\prime p)R_{ln}(u^\prime) = -\frac{1}{2\pi\eta_{ln}} \int_0^{1/2}du^\prime u^\prime R_{ln}(u^\prime) G_l(u,u^\prime).\label{Step2}
\end{equation}
The radial part of the PWFs can be further expanded as
\begin{equation}
R_{ln}(u) = (2u)^{|l|}\sum_{m^\prime} a^{ln}_{m^\prime}P^{(|l|,0)}_{m^\prime}(1 - 8u^2), \label{PWFDisk}
\end{equation}
where $a^{ln}_{m^\prime}$ are to be determined and $P^{(\alpha,\beta)}_m(x)$ are the Jacobi Polynomials. Multiplying both sides by $(2u)^{|l| + 1}P^{(|l|,0)}_{m}(1 - 8u^2)$, making the change of variables $x = 2u$, $x^\prime = 2u^\prime$, and $p \rightarrow 2p$, and using the relations~\cite{gradshteyn2014} 
\begin{align}
\int_0^1 dx \, x^{|l| + 1}P^{(|l|,0)}_m(1 - 2x^2) J_{|l|}(px) &= \frac{J_{|l| + 2m + 1}(p)}{p}, \label{Step4}
\\
\int_0^\infty dp \,\frac{J_{|l| + 2m + 1}(p)J_{|l| + 2m^\prime + 1}(p)}{p^2} &= \frac{(-1)^{m - m^\prime + 1}}{\pi[4(m - m^\prime)^2 -1][|l| + m + m^\prime + 1/2][|l| + m + m^\prime + 3/2]},
\end{align}
allows us to immediately solve the left-hand side (LHS) of equation \eqref{Step2} once we integrate in $x$. We have
\begin{equation}
\mbox{LHS} = \frac{1}{2}\sum_{m^\prime} \mathbb{K}^l_{mm^\prime} a^{ln}_{m^\prime},
\label{LHS}
\end{equation}
where 
\begin{equation}
\mathbb{K}^l_{mm^\prime} = \frac{(-1)^{m - m^\prime + 1}}{\pi[4(m -  m^\prime)^2 -1](|l| + m +  m^\prime + 1/2)(|l| + m +  m^\prime + 3/2)}, \,\,\,\, m,m^\prime = 0,1,2,3...
\end{equation}
The right-hand side (RHS) of equation \eqref{Step2} can be solved in the same way once we plug the expression for $G_l(u,u^\prime)$ and use the orthogonality relation~\cite{gradshteyn2014} 
\begin{equation}
\int_0^1 dx x^{2|l| + 1}P_i^{(|l|,0)}(1 - 2x^2)P_j^{(|l|,0)}(1 - 2x^2) = \frac{\delta_{ij}}{2(|l| + 2j + 1)}. \label{Step5}
\end{equation}
For $l \neq 0$, $G_l(u,u^\prime) = \frac{1}{2|l|}[(uu^\prime)^{|l|} + (u_</u_>)^{|l|}]$, where $u_> = \mbox{max}(u,u^\prime)$ and $u_< = \mbox{min}(u,u^\prime)$. Therefore, after integration over $x$, we obtain
\begin{equation}
\mbox{RHS} = -\frac{1}{8\pi\eta_{ln}}\sum_{m^\prime} \mathbb{G}^l_{mm^\prime} a^{ln}_{m^\prime},
\label{RHS}
\end{equation}
where
\begin{align}
\mathbb{G}^l_{mm^\prime} &= \frac{\delta_{m0}\delta_{m^\prime0}}{8|l|(|l| + 1)^2} + \frac{\delta_{mm^\prime}}{4(|l| + 2m^\prime)(|l| + 2m^\prime + 1)(|l| + 2m^\prime + 2)}  +\frac{\delta_{m+ 1,m^\prime}}{8(|l| + 2m+ 1)(|l| + 2m + 2)(|l| + 2m + 3)} \cr &+ \frac{\delta_{m,m^\prime+ 1}}{8(|l| + 2m^\prime+ 1)(|l| + 2m^\prime + 2)(|l| + 2m^\prime + 3)}, \,\,\,\, m,m^\prime = 0,1,2,3...
\end{align}
For $l = 0$, $G_l(x,x^\prime) = -\mbox{ln}(x_>)$ and the calculations are not as straightforward~\cite{fetter1986}. The result is the same as for $l \neq 0$, but the matrix $\mathbb{G}^l$ does not have the first term ($m, m^\prime \neq 0$) of the RHS of the previous equation. Finally, by combining Eqs. \eqref{LHS} and \eqref{RHS}, we obtain an eigenvalue equation for the vector ${\bf a}^{ln}  = \{a^{ln}_{m} \}$, 
\begin{equation}
\mathbb{G}^l{\bf a}^{ln} = -4\pi\eta_{ln}\mathbb{K}^l{\bf a}^{ln}\, .
\end{equation}
We solved this eigenvalue equation numerically for up to $m, m^\prime = 300$, obtaining satisfactory convergence. The normalization of ${\bf a}^{ln}$ is obtained by enforcing Eq. \eqref{PWForthogonality} to be satisfied, resulting in a normalization factor given by $\sqrt{8/\pi {\bf a}^{ln}\mathbb{G}^l{\bf a}^{ln}}$.

Finally, several of the results we demonstrated before for the TPSE admit simple semi-analytical expressions in the case of the nanodisk. In particular,
\begin{equation}
{\bf F}_{ln}({\bf R}_e) = \pi D\nabla_{{\bf R}_e}e^{il\phi_e}\sum_m a^{ln}_m\int_0^\infty \frac{dp}{p}e^{-2pz_e/D}J_{|l|}\left(\frac{2r_e p}{D}\right)J_{|l| + 2m + 1}(p).
\end{equation}
For the case analyzed in the main text of a dipole placed on the symmetry axis of the nanodisk,  the integral above can be solved analytically, resulting in
\begin{align}
F_{ln,x}(z_e) &= \pi\delta_{|l|1}\sum_{m = 0}^\infty a^{1n}_m\frac{\left(\sqrt{\frac{4z_e^2}{D^2} + 1} - \frac{2z_e}{D}\right)^{2(m + 1)}}{\sqrt{\frac{4z_e^2}{D^2} + 1}} = -iF_{ln,y}(z_e) , \cr
F_{ln,z}(z_e) &= -2\pi\delta_{l0}\sum_{m = 1}^\infty a^{0n}_m\frac{\left(\sqrt{\frac{4z_e^2}{D^2} + 1} - \frac{2z_e}{D}\right)^{2m + 1}}{\sqrt{\frac{4z_e^2}{D^2} + 1}}.
\end{align}
Also,
\begin{equation}
\bm{\zeta}_{ln} := \frac{\pi}{32}\delta_{|l|1}a^{1n}_0({\bf \hat{x}} + \mbox{sgn}(l)i{\bf \hat{y}})
\end{equation}
and
\begin{align}
{\bf d}_{a,ind}(z_e,\omega) =\frac{\pi}{16}\sum_n \frac{a^{1n}_0F_{1n,x}(z_e)}{1/\eta_{1n} - 1/\eta(\omega)} [{\bf d}_a - ({\bf d}_a\cdot{\bf \hat{z}}){\bf \hat{z}}] ,
\end{align}
which gives us a straightforward way to compute the radiative and non-radiative contributions for the spontaneous emission rate in this situation. We point that it is clear from these expressions that only the dipole modes ($l = 1$) contribute to the $x$ and $y$ Purcell factors and only the dark modes ($ l = 0$) contribute to the $z$ Purcell factor, which has a radiative part equal to  $1$.

%%%%%%%%%%%%%%%%%%%%%%%%%%%%%%%%%%%%%%%%%%%%%%%%%%%%%%%%%%%%%%%%%%%%%%%%%
%


\begin{thebibliography}{99}
 
 \bibitem{haroche2013} Haroche, S. Nobel Lecture: Controlling photons in a box and exploring the quantum to classical boundary. {\em Rev. Mod. Phys.} {\bf 85}, 1083-1102 (2013).

 \bibitem{aharonovich2016} Aharonovich, I., Englund, D. \& Toth, M. Solid-state single-photon emitters. {\em Nat. Photonics} {\bf 10}, 631-641 (2016).
 
 \bibitem{kwiat95} P. G. Kwiat {\em et al}. New high-intensity source of polarization-entangled photon pairs. {\em Phys. Rev. Lett.} {\bf 75}, 4337 (1995).
 
\bibitem{goppert1931}
G{\"o}ppert-Mayer, M.  {\"U}ber Elementarakte mit zwei Quantenspr{\"u}ngen.
 {\em Annalen der Physik} {\bf 401}, 273-294 (1931).

 \bibitem{zalialiutdinov2018} Zalialiutdinov, T. A., Solovyec, D. A., Labzowsky, L. N., \& Plunien, G. QED theory of multiphoton transitions in atoms and ions. {\em Phys. Rep.} {\bf 737}, 1-84 (2018).

\bibitem{lipeles1965}
Lipeles, M., Novick, R. \& Tolk, N. Direct detection of two-photon emission from the
metastable state of singly ionized helium. {\em Phys. Rev. Lett.} {\bf 15}, 690 (1965).

\bibitem{bannett1982}
Bannett, Y. \&  Freund, I. Two-photon x-ray emission from inner-shell transitions. {\em Phys. Rev. Lett.} {\bf 49}, 539 (1982).

\bibitem{cesar1996} Cesar,  C. L., et al. Two-photon spectroscopy of trapped atomic hydrogen. 
{\em Phys. Rev. Lett.} {\bf 77}, 255 (1996).

\bibitem{hayat2008}
Hayat, A., Ginzburg, P. \& Orenstein, M. Observation of two-photon emission from semiconductors. {\em Nat. Photonics} {\bf 2}, 238-241 (2008).
 
 \bibitem{ota2011}
Ota, Y., Iwamoto, S., Kumagai, N. \& Arakawa, Y. Spontaneous two-photon emission from a single quantum dot. {\em Phys. Rev. Lett.} {\bf 107}, 233602 (2011).

\bibitem{wang2019} H. Wang {\em et al}. On-demand semiconductor
source of entangled photons which simultaneously has high fidelity, efficiency, and
indistinguishability. {\em Phys. Rev. Lett.} {\bf 122}, 113602 (2019).

\bibitem{Tame2013}   
Tame, M. S., et al. Quantum plasmonics. {\em Nat. Phys.} {\bf 9}, 329-340 (2013).
 
\bibitem{lodahl2015} 
Lodahl, P., Mahmoodian, S. \& Stobbe, S. Interfacing single photons and single quantum dots with photonic nanostructures. {\em Rev. Mod. Phys.} {\bf 87}, 347-400 (2015).

\bibitem{poddubny2012} 
Poddubny, A. N., Ginzburg, P., Belov, P. A., Zayats, A. V. \& Kivshar, Y. S. Tailoring and enhancing spontaneous two-photon emission using resonant plasmonic nanostrucutres. {\em Phys. Rev. A} {\bf  86}, 033826 (2012).

\bibitem{nevet2010}
Nevet, A., et al. Plasmonic nanoantennas for broad-band enhancement of
two-photon emission from semiconductors. {\em Nano Lett.} {\bf 10}, 1848-1852 (2010).

\bibitem{goncalves2020}
 Gon{\c{c}}alves, P. A. D., et al. Plasmon--emitter interactions at the nanoscale. {\em Nat. Comm.} {\bf 11}, 366 (2020).

\bibitem{rivera2016}
Rivera, N., Kaminer, I., Zhen, B., Joannopoulos, J. D. \& Soljacic, M.
Shrinking light to allow forbidden transitions on the atomic scale. {\em Science} {\bf 353}, 263-269 (2016).

\bibitem{rivera2017}
Rivera, N., Rosolen, G., Joannopoulos, J. D., Kaminer, I. \& Soljacic, M.
Making two-photon processes dominate one-photon processes using mid-IR phonon polaritons.
{\em Proc. Natl. Acad. Sci.} {\bf 114}, 13607-13612 (2017).

\bibitem{hoang2015} Hoang, T. B. et al. Ultrafast spontaneous emission source using plasmonic nanoantennas. {\em Nat. Commun.} {\bf 6}, 778 (2015).

\bibitem{note1}  We interchangeably use the  acronym TPSE and the terminology ``two-quanta spontaneous emission" to denote any {\it two-quanta} emissions, regardless of their radiative or non-radiative nature. Whenever this may lead to confusion, we explicitly identify the emission channel.

\bibitem{maniyara2019} 
Maniyara, R. A., et al. 
Tunable plasmons in ultrathin metal films. {\em Nat. Photonics} {\bf 13}, 328-333 (2019).

\bibitem{el-fattah2019}
Abd El-Fattah, Z. M., et al.
Plasmonics in atomically thin crystalline silver films. {\em ACS Nano} {\bf 13}, 7771-7779 (2019).

\bibitem{lodhal2004} Lodhal, P. et al. Controlling the dynamics of spontaneous emission from quantum dots by photonic crystals. {\em Nature} {\bf 430}, 654 (2004).

\bibitem{muniz2019} 
Muniz, Y., Szilard, D., Kort-Kamp, W. J. M., da Rosa, F. S. S. \& Farina, C.
Quantum two-photon emission in a photonic cavity. {\em Phys. Rev. A} {\bf 100}, 023818 (2019).

\bibitem{SI} See supplemental material at [url] for detailed derivations of the results presented here, which also includes Ref.  26-29.

\bibitem{sakurai2014} Sakurai, J. J., Napolitano, J. J. Modern quantum mechanics. Pearson Higher Ed (2014).

\bibitem{buhmann2012} Buhmann, S. Y. Dispersion Forces I: Macroscopic Quantum Electrodynamics. Springer (2012)

\bibitem{gradshteyn2014} Gradshteyn,I. S.,Ryzhik, I. M. Table of integrals, series, and products. Academic press (2014)

\bibitem{craig84} Craig, D. P., Thirunamachandran, T. Molecular quantum electrodynamics: an introduction to radiation-molecule interactions. Academic
Press (1984)

\bibitem{novotny2012}
Novotny, L. \& Hecht, B. Principles of Nano-Optics (Cambridge University Press, Cambridge, 2012).

\bibitem{note2}  We consider the case where the TPSE rate can be expressed in a basis of unit vectors $\{{\bf \hat{e}}_1,{\bf \hat{e}}_2,{\bf \hat{e}}_3\}$ that simultaneously diagonalize the imaginary part of the electromagnetic Green's function at frequencies $\omega$ and $\omega_t-\omega$, which simplifies the expression for the decay rate as in Eq. \eqref{GammaTotal}. Our key findings are valid beyond this special case, and the general expression can be found in the supporting information.

\bibitem{milonni2013}Milonni, P. W. The quantum vacuum: an introduction to quantum electrodynamics. Academic press (2013).

\bibitem{carminati2015} 
Carminati, R., et al. Electromagnetic density of states in complex plasmonic systems.
{\em Surf. Sci. Rep.} {\bf 70}, 1-41 (2015).

\bibitem{anger2006}  Anger, P.,  Bharadwaj, P., and  Novotny, L. Enhancement and Quenching of Single-Molecule Fluorescence. {\em Phys. Rev. Lett.} {\bf 96}, 113002 (2006).

\bibitem{manjavacas2015} 
Garc\'ia de Abajo, F. J. \& Manjavacas, A.    Plasmonics in atomically thin materials. 
{\em Faraday Discuss.} {\bf 178}, 87-1073548 (2015).

\bibitem{yu2017} 
Yu, R., Cox, J. D. Saavedra, J. R. M. \&  Garcia de Abajo, F. J.
Analytical modeling of graphene plasmons. {\em ACS Photonics} {\bf 4}, 3106-3114 (2017).

\bibitem{novotny2007} Bharadwaj, P., Novotny, L. Spectral dependence of single molecule fluorescence enhancement. Optics Express 15, 14266-14274 (2007)

\bibitem{miroshnichenko2010} 
Miroshnichenko, A. E., Flach, S. \& Kivshar, Y. S.
Fano resonances in nanoscale structures. {\em Rev. Mod. Phys.} {\bf 82}, 2257 (2010).

\bibitem{lukyanchuk2010} 
Luk'yanchuk, B., et al. 
The Fano resonance in plamonic nanostructures and metamaterials. {\em Nat. Mater.} {\bf 9}, 707-715 (2010).

\bibitem{limonov2017} 
Limonov, M. F., Rybin, M. V. Poddubny, A. N. \& Kivshar, Y. S. 
Fano resonances in photonics. {\em Nat. Photonics} {\bf 11}, 543-554 (2017).

\bibitem{manjavacas2014} 
Manjavacas, A. \& Garc\'ia de Abajo, F. J. Tunable plasmons in atomically thin gold nanodisks. 
{\em Nat. Commun.} {\bf 5}, 3548 (2014).

\bibitem{fetter1986} 
Fetter, A. L. 
Magnetoplasmons in a two-dimensional electron fluid: Disk geometry. {\em Phys. Rev. B} {\bf  33}, 5221 (1986).

\bibitem{tamagnone2018} Tamagnone, M., et al. 
Ultra-confined mid-infrared resonant phonon polaritons in van der Waals nanostructures. {\em Science Advances} {\bf 4}, eaat7189 (2018).

\bibitem{neto09} Neto, A. H. C., Guinea, F., Peres, N. M. R., Novoselov, K. S., Geim, A. K. The electronic properties of graphene
{\em Rev. Mod. Phys.} {\bf 81}, 109 (2009).

\bibitem{bolotin2008} Bolotin, K. I., et al. 
Ultrahigh electron mobility in suspended graphene. {\em Solid State Commun.}  {\bf 146}, 351-355 (2008).

\bibitem{dean2010} Dean, C. R., et al.
 Boron nitride substrates for high-quality graphene electronics. {\em Nat. Nanotechnol.} {\bf 5}, 722-726 (2010).

\bibitem{note3} The TPSE $\gtrsim 10^4$ enhancement accounts for spectral shifts in the plasmonic resonances and screening of the Purcell factors due to the presence of the SiO$_2$ substrate. We estimate that off-axis quantum dots will reduce the enhancement by about one order of magnitude due to weaker coupling to the resonant mode.

\bibitem{gao2010} Gao, W.-B. et al. Experimental demonstration of a hyper-entangled ten-qubit Schr\"odinger cat state. {\em Nature Physics} {\bf 6}, 331-335 (2010).

\end{thebibliography}

\begin{thebibliography}{99}
 
 %%Review
 \bibitem{milonni2013}
Milonni, P. W. The quantum vacuum: an introduction to quantum electrodynamics. {\em Academic press} (2013).

 \bibitem{sakurai2014}
Sakurai, J. J., Napolitano, J. J. Modern quantum mechanics. {\em Pearson Higher Ed} (2014).

 \bibitem{thiru1984}
Craig, D. P., Thirunamachandran, T. Molecular quantum electrodynamics: an introduction to radiation-molecule interactions. {\em Academic Press} (1984).
	 
 \bibitem{Muniz2019}
Muniz, Y., Szilard, D., Kort-Kamp, W. J. M., da Rosa, F. S. S. \& Farina, C. Quantum two-photon emission in a photonic cavity. {\em Phys. Rev. A} {\bf 100}, 023818 (2019).

 \bibitem{novotny2012}
Novotny, L., Hecht, B. Principles of nano-optic. {\em Cambridge university press} (2012)
 
 \bibitem{buhmann2012}
Buhmann, S. Y. Dispersion Forces I: Macroscopic Quantum Electrodynamics. {\em Springer} (2012)

 \bibitem{abajo2015}
Garcia de Abajo, F. J., Manjavacas, A. Plasmonics in atomically thin materials. {\em Faraday discussions} {\bf 178}, 87-107 (2015)

 \bibitem{yu2017}
Yu, R., Cox, J. D. Saavedra, J. R. M. \&  Garcia de Abajo, F. J. Analytical modeling of graphene plasmons. {\em ACS Photonics} {\bf 4}, 3106-3114 (2017).

 \bibitem{novotny2007}
Bharadwaj, P., Novotny, L. Spectral dependence of single molecule fluorescence enhancement.  {\em Optics Express} {\bf 15}, 14266-14274 (2007)

 \bibitem{fetter1986}
Fetter, A. L. Magnetoplasmons in a two-dimensional electron fluid: Disk geometry. {\em Phys. Rev. B} {\bf  33}, 5221 (1986).

 \bibitem{gradshteyn2014}
Gradshteyn,I. S.,Ryzhik, I. M. Table of integrals, series, and products. {\em Academic press} (2014)
  
\end{thebibliography}
\end{document}